\documentclass[%
 aip,
 amsmath,amssymb,
 reprint,%
]{revtex4-1}

\usepackage{graphicx}
\usepackage{dcolumn}
\usepackage{bm}
\usepackage[colorlinks,linkcolor=blue,anchorcolor=blue,citecolor=blue]{hyperref}

\usepackage[utf8]{inputenc}
\usepackage[T1]{fontenc}
\usepackage{mathptmx}

\begin{document}

\preprint{AIP/123-QED}

\title{Deep-learning-enabled geometric constraints and phase unwrapping for single-shot absolute 3D shape measurement}

\author{Jiaming Qian}
\email{jiaming_qian@njust.edu.cn.}
 \affiliation{School of Electronic and Optical Engineering, Nanjing University of Science and Technology, No. 200 Xiaolingwei Street, Nanjing, Jiangsu Province 210094, China.}
 \affiliation{Jiangsu Key Laboratory of Spectral Imaging \& Intelligent Sense, Nanjing University of Science and Technology, Nanjing, Jiangsu Province 210094, China.}
 \affiliation{Smart Computational Imaging (SCI) Laboratory, Nanjing University of Science and Technology, Nanjing,Jiangsu Province 210094, China.}

 \author{Shijie Feng}
 \email{ShijieFeng@njust.edu.cn.}
 \affiliation{School of Electronic and Optical Engineering, Nanjing University of Science and Technology, No. 200 Xiaolingwei Street, Nanjing, Jiangsu Province 210094, China.}
 \affiliation{Jiangsu Key Laboratory of Spectral Imaging \& Intelligent Sense, Nanjing University of Science and Technology, Nanjing, Jiangsu Province 210094, China.}
 \affiliation{Smart Computational Imaging (SCI) Laboratory, Nanjing University of Science and Technology, Nanjing,Jiangsu Province 210094, China.}

 \author{Tianyang Tao}
 \affiliation{School of Electronic and Optical Engineering, Nanjing University of Science and Technology, No. 200 Xiaolingwei Street, Nanjing, Jiangsu Province 210094, China.}
 \affiliation{Jiangsu Key Laboratory of Spectral Imaging \& Intelligent Sense, Nanjing University of Science and Technology, Nanjing, Jiangsu Province 210094, China.}
 \affiliation{Smart Computational Imaging (SCI) Laboratory, Nanjing University of Science and Technology, Nanjing,Jiangsu Province 210094, China.}

 \author{Yan Hu}
 \affiliation{School of Electronic and Optical Engineering, Nanjing University of Science and Technology, No. 200 Xiaolingwei Street, Nanjing, Jiangsu Province 210094, China.}
 \affiliation{Jiangsu Key Laboratory of Spectral Imaging \& Intelligent Sense, Nanjing University of Science and Technology, Nanjing, Jiangsu Province 210094, China.}
 \affiliation{Smart Computational Imaging (SCI) Laboratory, Nanjing University of Science and Technology, Nanjing,Jiangsu Province 210094, China.}

 \author{Yixuan Li}
 \affiliation{School of Electronic and Optical Engineering, Nanjing University of Science and Technology, No. 200 Xiaolingwei Street, Nanjing, Jiangsu Province 210094, China.}
 \affiliation{Jiangsu Key Laboratory of Spectral Imaging \& Intelligent Sense, Nanjing University of Science and Technology, Nanjing, Jiangsu Province 210094, China.}
 \affiliation{Smart Computational Imaging (SCI) Laboratory, Nanjing University of Science and Technology, Nanjing,Jiangsu Province 210094, China.}

 \author{Qian Chen}
 \email{chenqian@njust.edu.cn.}
 \affiliation{School of Electronic and Optical Engineering, Nanjing University of Science and Technology, No. 200 Xiaolingwei Street, Nanjing, Jiangsu Province 210094, China.}
 \affiliation{Jiangsu Key Laboratory of Spectral Imaging \& Intelligent Sense, Nanjing University of Science and Technology, Nanjing, Jiangsu Province 210094, China.}

 \author{Chao Zuo}
 \email{ zuochao@njust.edu.cn.}
 \affiliation{School of Electronic and Optical Engineering, Nanjing University of Science and Technology, No. 200 Xiaolingwei Street, Nanjing, Jiangsu Province 210094, China.}
 \affiliation{Jiangsu Key Laboratory of Spectral Imaging \& Intelligent Sense, Nanjing University of Science and Technology, Nanjing, Jiangsu Province 210094, China.}
 \affiliation{Smart Computational Imaging (SCI) Laboratory, Nanjing University of Science and Technology, Nanjing,Jiangsu Province 210094, China.}

\date{\today}

\begin{abstract}
  Fringe projection profilometry (FPP) has becoming more prevalently adopted in intelligent manufacturing, defect detection and some other important applications. In FPP, how to efficiently recover the absolute phase has always been a great challenge. The stereo phase unwrapping (SPU) technologies based on geometric constraints can eliminate phase ambiguity without projecting any additional patterns, which maximizes the efficiency of the retrieval of absolute phase. Inspired by recent successes of deep learning for phase analysis, we demonstrate that deep learning can be an effective tool that organically unifies phase retrieval, geometric constraints, and phase unwrapping into a comprehensive framework. Driven by extensive training dataset, the neural network can gradually "learn" how to transfer one high-frequency fringe pattern into the "physically meaningful", and "most likely" absolute phase, instead of "step by step" as in conventional approaches. Based on the properly trained framework, high-quality phase retrieval and robust phase ambiguity removal can be achieved based on only single-frame projection. Experimental results demonstrate that compared with traditional SPU, our method can more efficiently and stably unwrap the phase of dense fringe images in a larger measurement volume with fewer camera views. Limitations  about the proposed approach are also discussed. We believe the proposed approach represents an important step forward in high-speed, high-accuracy, motion-artifacts-free absolute 3D shape measurement for complicated object from a single fringe pattern.
\end{abstract}

\maketitle

\section{\label{sec:level1}Introduction}
Optical non-contact three-dimensional (3D) shape measurement techniques have been widely applied for many aspects, such as intelligent manufacturing, reverse engineering, heritage digitalization and so on \cite{salvi2010state}. The fringe projection profilometry (FPP) \cite{gorthi2010fringe} is one of the most popular optical 3D imaging techniques due to its simple hardware configuration, flexibility in implementation, and high measurement accuracy. 

With the development of imaging and projection devices, it becomes possible to realize high speed 3D shape measurement based on FPP \cite{zhang2010superfast,zuo2013high,tao2018high,feng2018robust,feng2018high}. Meanwhile, the acquisition of high-quality 3D information in high-speed scenarios is increasingly crucial to many applications, such as online quality inspection, stress deformation analysis, rapid reverse molding, etc \cite{zuo2018micro,qian2019high}. To achieve 3D measurement in high-speed scenarios, efforts are usually carried out by reducing the number of images required for per reconstruction to improve measurement efficiency. The ideal way is to obtain 3D data in a single frame. Recently, we have realized high-accuracy phase acquisition from a single fringe pattern by using deep learning \cite{feng2019fringe,feng2019micro}. However, these works just obtain single-shot wrapped phase. To realize 3D measurement, phase unwrapping is required, which is one of the operations in FPP that most affect measurement efficiency. The most commonly used phase unwrapping methods are temporal phase unwrapping (TPU) algorithms \cite{zuo2016temporal,zhang2006time}, which recover the absolute phase with the assistance of Gray-code patterns or multi-wavelength fringes. However, the requirement of additional patterns decreases the measurement efficiency. The stereo phase unwrapping (SPU) \cite{weise2007fast} method based on geometric constraints can solve the phase ambiguity problem through the spatial relationships between multiple cameras and one projector without projecting any auxiliary patterns. Although requiring more cameras (at least two) than traditional methods, SPU indeed maximizes the efficiency of FPP. However, conventional SPU is generally insufficient to robustly unwrap the phase of dense fringe images, while increasing the frequency of fringes is essential to the measurement accuracy. To solve this trade-off, some auxiliary algorithms are proposed, which usually focus on four directions. (1) The first directions utilize spatial phase unwrapping methods \cite{su2004reliability} to reduce phase unwrapping errors of SPU \cite{weise2007fast,garcia2012consistent}. As the disadvantages of spatial phase unwrapping, these methods cannot handle discontinuous or disjoined phases. (2) The second directions enhance the robustness of SPU by embedding auxiliary information in the fringe patterns \cite{lohry2014high, tao2016real}. Since assistance based on intensity information is provided, the sensitivity of intensity to ambient light noise and large surface reflectivity variations of objects will cause them to fail. (3) The third aspect is to increase the number of perspectives and recover the absolute phase through more geometric constraints \cite{tao2017high}. This method is more adaptive for complex scene measurement, but comes at the cost of increased cost. Besides, simply increasing the number of views is insufficient to unwrap the phase of dense fringe images, which needs to be combined with (4) the depth constraint strategy \cite{brauer2011using,li2013multiview,liu2017high}. However, conventional depth constraint strategy can only unwrap the phase in a narrow depth range, and how to set a suitable depth constraint range is also difficult. The adaptive depth constraint (ADC) \cite{tao2018high,qian2019motion} strategy can enlarge the measurement volume and automatically select the depth constraint range, but only if the correct absolute phase can be obtained for the first measurement. In addition, since the stability of SPU relies on the similarity of the phase information of matching points in different perspectives \cite{tao2017high}, on the one hand, SPU requires high-quality system calibration and is more difficult to implement algorithmically than other phase unwrapping methods such as TPU; on the other hand, it has high demands for the quality of the wrapped phase. So that the wrapped phase in SPU is usually acquired by phase-shifting (PS) algorithm \cite{zuo2018phase}, which is a multi-frame phase acquisition method with high spatial resolution and high measurement accuracy. However, the use of multiple fringe patterns reduces the measurement efficiency of SPU. Another commonly used phase acquisition technologies are Fourier transform (FT) methods \cite{su2010dynamic,huang2010comparison} with single-shot nature, which are not suitable for SPU due to the poor imaging quality around discontinuities and isolated areas in the phase map. 

From above discussion, it is not difficult to know that although SPU is best suitable for 3D measurement in high-speed scenes, it still has some defects, such as limited measurement volume, inability to robustly achieve phase unwrapping of high-frequency fringe images, loss of measurement efficiency due to reliance on multi-frame phase acquisition methods, complexity of algorithm implementation and so on. Inspired by successes of deep learning in FPP \cite{van2019deep,feng2019fringe,feng2019micro,yin2019temporal} and the advance of geometric constraints, on the basis of our previous deep-learning-based works, we further push deep learning into phase unwrapping and incorporate geometric constraints into the neural network. In our work, geometric constraints are implicit in the neural network rather than directly using calibration parameters, which simplifies the entire process of phase unwrapping and avoids the complex adjustment of various parameters. With extensive data training, the network can "learn" to obtain the "physically meaningful" absolute phase from single-frame projection without the conventional "step-by-step" calculation. Compared with traditional SPU, our approach more robustly unwraps the phase of the higher frequency with fewer perspectives in a larger range. In addition, the limitations of proposed approach are also analyzed in the conclusions and discussions section.

\section{Principle}
\subsection{Phase retrieval and unwrapping with PS and SPU}\label{section:2.1}
\begin{figure}
	\includegraphics[width=1\linewidth]{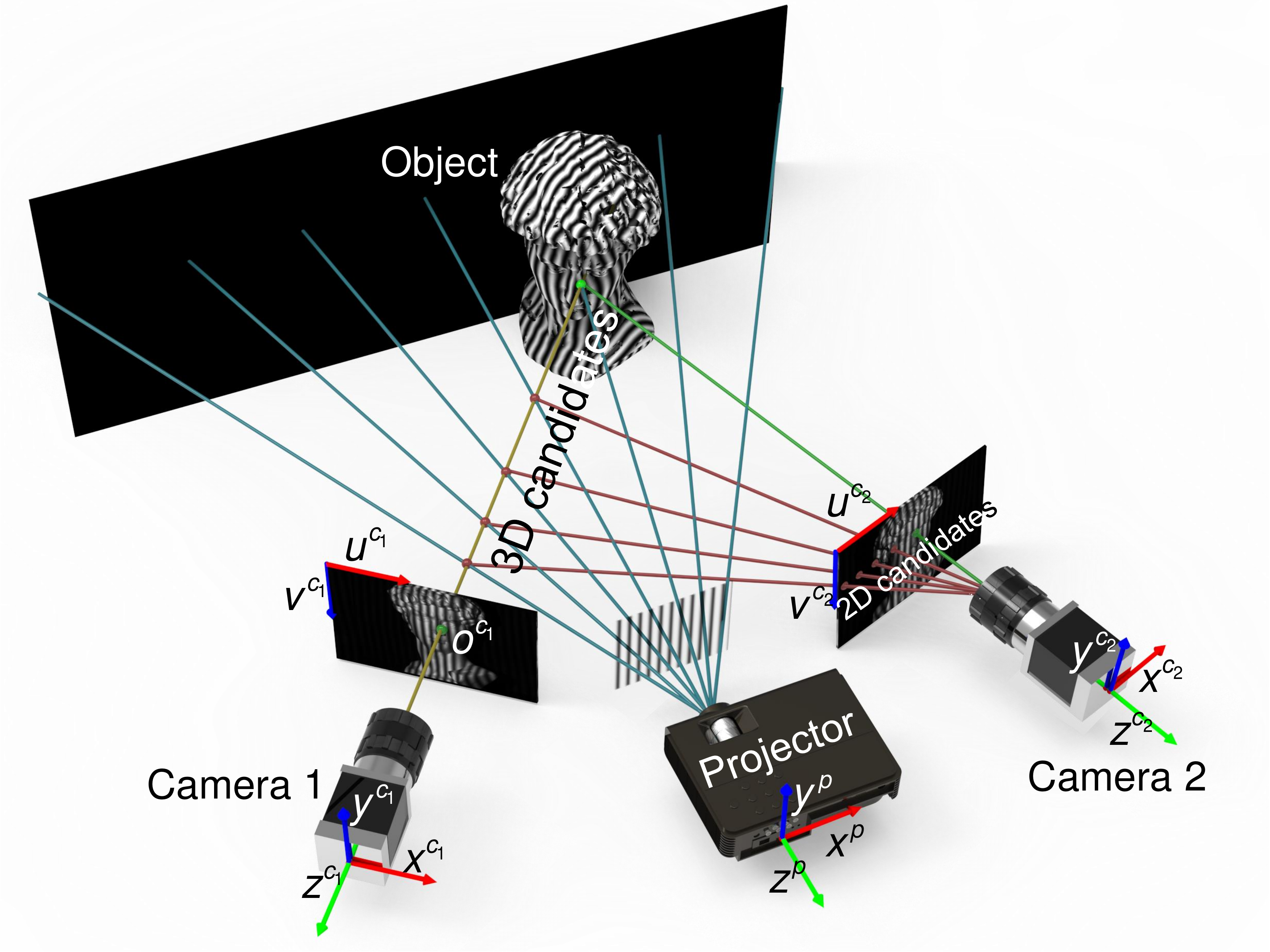}
	\caption{The principle of SPU.}
	\label{fig1}
\end{figure}

As shown in Fig. \ref{fig1}, a typical SPU-based system consists of one projector and two cameras. The fringe images are projected by the projector, and then modulated by the object, and finally captured by two cameras. For $N$-step PS algorithm, the fringe patterns captured by Camera 1 can be expressed as:
\begin{equation}
  I_n({u^c},{v^c}) = {A}({u^c},{v^c}) + {B}({u^c},{v^c})\cos ({\Phi }({u^c},{v^c}) + 2\pi n/N),
  \label{eq1}
  \end{equation}
  where $I_n$ represents the $(n+1)$th captured image, $n = 0,1,... ,N - 1$,  $({u^c},{v^c})$ is the camera pixel coordinate, $A$  is the average intensity map, $B$  is the amplitude intensity map, ${\Phi}$ is the absolute phase map, and $2\pi n/N$ is the phase shift. With the least square method \cite{hu2019microscopic}, the wrapped phase ${\phi}$ can be obtained:
\begin{equation}
	{\phi} = \arctan \frac{{{M}}}{{{D}}}{\rm{ = }}\arctan \frac{{\sum\nolimits_{n = 0}^{N - 1} {I_n\sin (2\pi n/N)} }}{{\sum\nolimits_{n = 0}^{N - 1} {I_n\cos (2\pi n/N)} }},
	\label{eq2}
\end{equation}
where $({u^c},{v^c})$ is omitted for convenience, and $M$ and $D$ represent the numerator and denominator of arctangent function respectively. The absolute and wrapped phases satisfy the following relation:
\begin{equation}
	{\Phi} = {\phi } + 2{k}\pi,
\label{eq5}
\end{equation}
where $k$ is the fringe order, ${k} \in [0, K - 1]$, and $K$ denotes the number of the used 
\begin{figure*}
	\centering
	\includegraphics[width=1\linewidth]{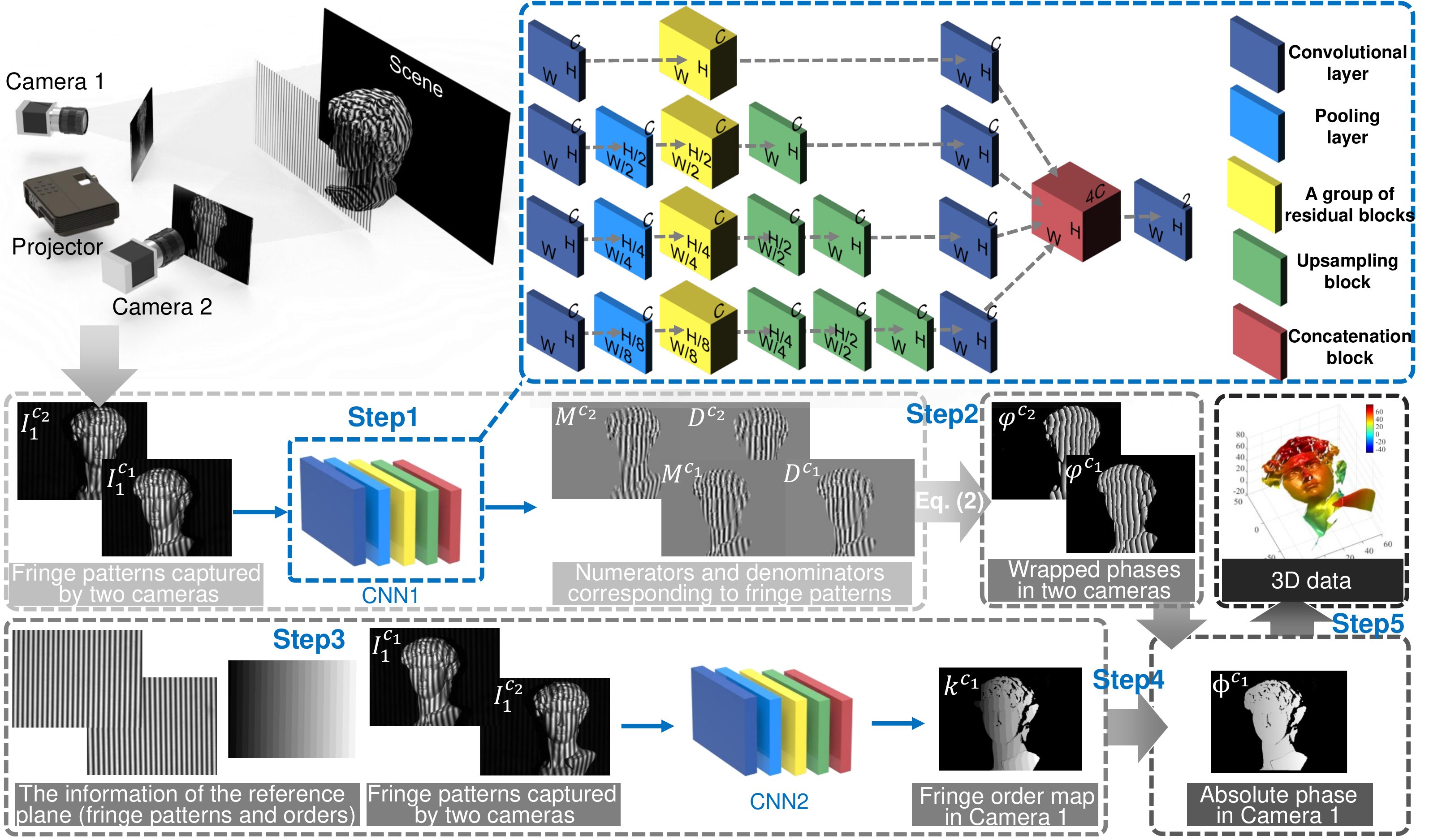}
	\caption{The flowchart of our method.}
	\label{fig2}
	\end{figure*}  
fringes. The fringe order $k$ can be obtained by using SPU based on geometric constraints. For an arbitrary point ${o^{{c_1}}}$ in Camera 1, it has $K$ possible fringe orders corresponding to $K$ absolute phases, with which $K$ 3D candidate points can be reconstructed by the calibration parameters between Camera 1 and the projector. The retrieved 3D candidates can be projected into Camera 2 to obtain its corresponding 2D candidates. Among these 2D candidates, there must be a correct matching point which has the more similar wrapped phase to ${o^{{c_1}}}$ than other candidates. With this feature, the matching point can be determined through the phase similarity check, and then the phase ambiguity of ${o^{{c_1}}}$ can be eliminated. However, due to calibration errors and ambient light interference, some wrong 2D candidates may have more similar phase value to ${o^{{c_1}}}$ than the correct matching point. Furthermore, the higher the frequency of the used fringes, the more candidates, the more likely such situation will happen. Therefore, in order to alleviate this issue, multi-step PS algorithm with higher measurement accuracy and robustness towards ambient illumination is preferred, and high-frequency fringe patterns are not recommended.

To enhance the stability of SPU, the common methods are to increase the number of views or apply the depth constraint strategy. The former, at the cost of increased hardware costs,  further projects 2D candidates of Camera 2 into the third or even the fourth camera for phase similarity check to exclude more wrong 2D candidates. The latter, at the cost of increased algorithm complexity, can eliminate some wrong 3D candidates outside the depth constraint range in advance. However, the conventional depth constraint algorithm is only effective in a narrow volume. Generally, the SPU with at least three cameras and assisted with ADC (the most advanced and complex depth constraint algorithm) can achieve robust phase unwrapping on the premise that the correct absolute phase is obtained for the first measurement \cite{tao2018high,qian2019motion}. However, complex systems and algorithms make such strategy difficult to implement.

\subsection{Phase retrieval and unwrapping with deep learning}\label{section:2.2}
The ideal SPU should be to use only two cameras and single frame projection to achieve robust phase unwrapping of dense fringe images in a large measurement volume without any complicated auxiliary algorithms. To this end, inspired by recent successes of deep learning techniques in phase analysis, we combine deep neural networks and SPU to develop a deep-learning-enabled geometric constraints and phase unwrapping method. The flowchart of our approach is shown in Fig. \ref{fig2}. We construct two four-path convolutional neural networks (CNN1 and CNN2) with the same structure (except for different inputs and outputs) to learn to obtain the high-quality phase information and unwrap the wrapped phase. The detailed architectures of the networks are provided in \textbf{Appendix A}.

\begin{figure*}
	\includegraphics[width=0.9\linewidth]{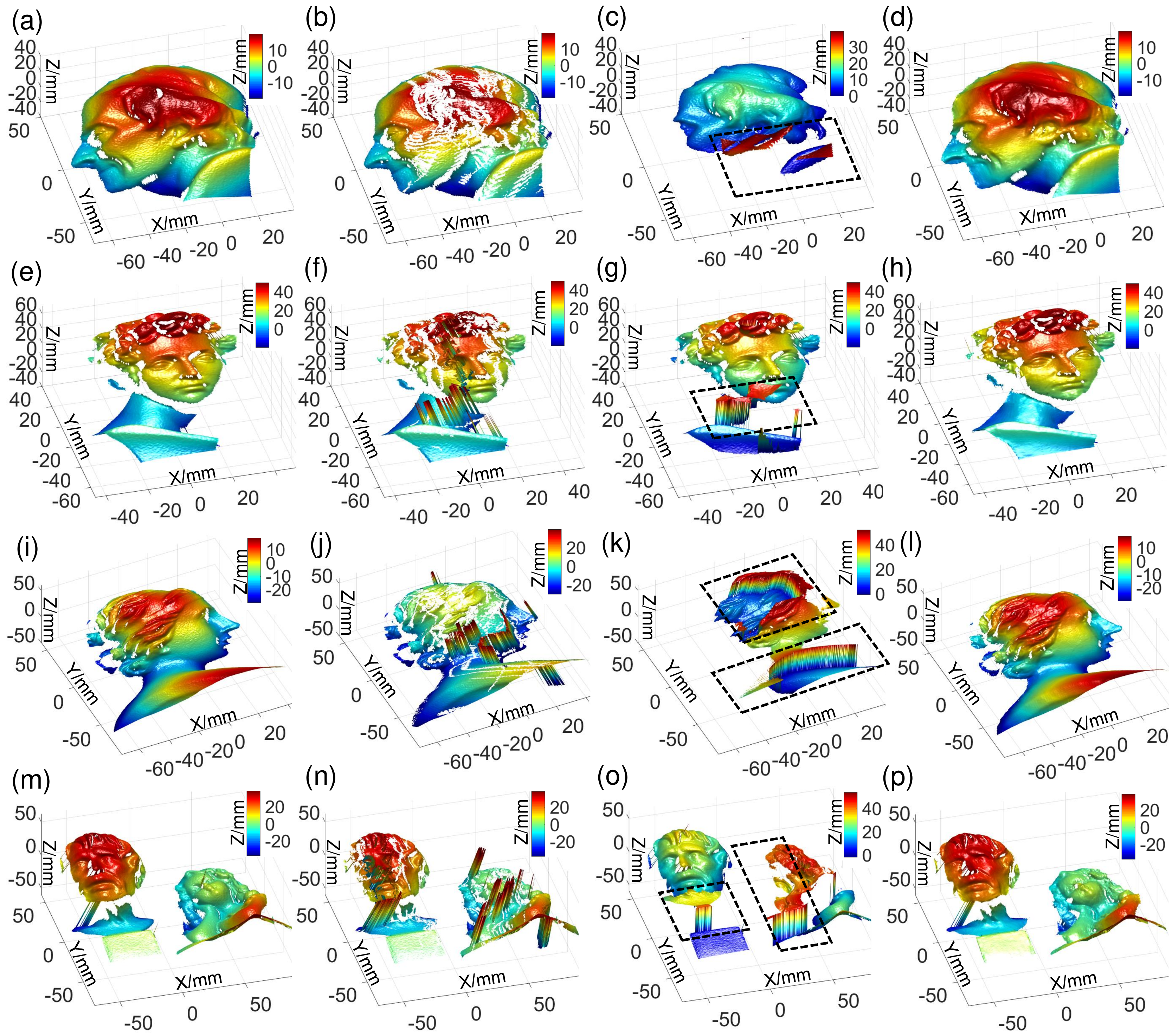}
	\caption{Measurement results of four static scenes. (a), (e), (i), (m) The results measured by the first method (taken as the ground-truth data). (b), (f), (j), (n) The results measured by the second method. (c), (g), (k), (o) The results measured by the third method. (d), (h), (l), (p) The results measured by our method.}
	\label{fig3}
\end{figure*}
\begin{figure*}
	\includegraphics[width=0.9\linewidth]{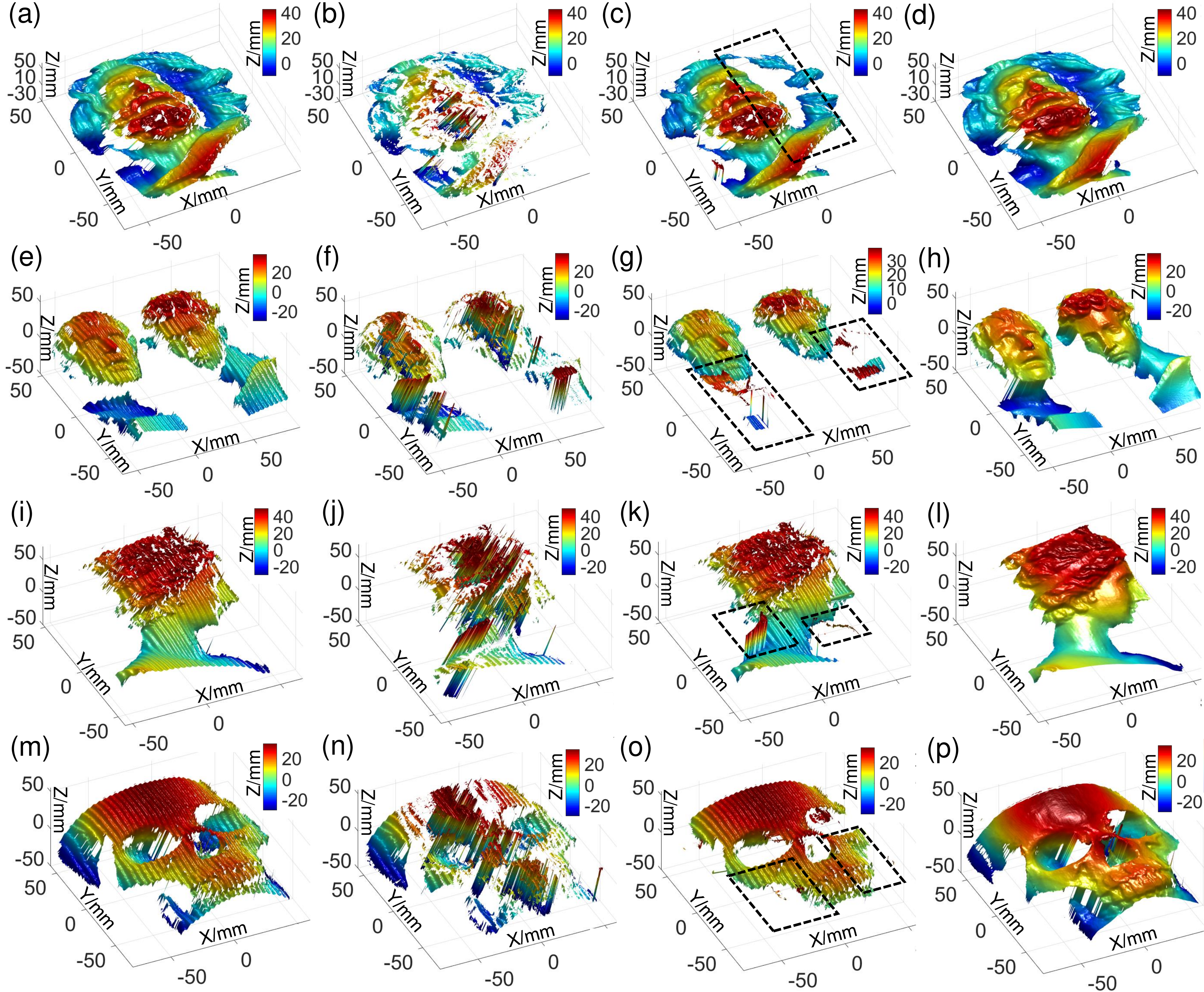}
	\caption{Measurement results of four dynamic scenes. (a), (e), (i), (m) The results measured by the first method. (b), (f), (j), (n) The results measured by the second method. (c), (g), (k), (o) The results measured by the third method. (d), (h), (l), (p) The results measured by our method. (Multimedia view: see \textcolor[rgb]{0,0,1}{Visualization 1}, \textcolor[rgb]{0,0,1}{Visualization 2}, \textcolor[rgb]{0,0,1}{Visualization 3} and \textcolor[rgb]{0,0,1}{Visualization 4} for the whole process of the first scene)}
	\label{fig4}
\end{figure*}
Next we will discuss our algorithm steps. $Step 1$: To achieve high-quality wrapped phase information retrieval, the physical model of conventional PS algorithm is considered. We separately input the single-frame fringe images captured by Camera 1 and Camera 2 into CNN1, and outputs are the numerators $M$ and denominators $D$ of the arctangent function corresponding to the two fringe patterns instead of directly linked wrapped phases, since such strategy bypasses the difficulties associated with reproducing abrupt $2\pi$ phase wraps to provide a high-quality phase estimate \cite{feng2019fringe}. $Step 2$: After predicting the numerator and denominator  terms, high-accuracy wrapped phase maps of Camera 1 and Camera 2 can be obtained according to Eq. \ref{eq2}. $Step 3$: To realize the phase unwrapping, enlightened by the geometry-constraint-based SPU described in Section \ref{section:2.1} which can remove phase ambiguity through spatial relationships between multiple perspectives, the fringe patterns of two perspectives are fed into CNN2. Meanwhile, we integrate the idea of assisting phase unwrapping with reference plane information \cite{an2016pixel} to our network, and add the data of a reference plane to the inputs to allow CNN2 to more effectively acquire the fringe orders of the measured object. Thus, the raw fringe patterns captured by two cameras, as well as the reference information (containing two fringe images of the reference plane captured by two cameras, and the fringe order map of the reference plane in the perspective of Camera 1) are fed into CNN2. It is worth mentioning that the reference plane information is obtained in advance and subsequent experiments do not need to obtain it repeatedly, which means there is just one extra reference information for the whole setup necessary. The output of CNN2 is the fringe order map of the measured object in Camera 1. $Step 4$: Through the wrapped phases and the fringe orders obtained by the previous steps, high-quality unwrapped phase can be recovered by Eq. \ref{eq5}. $Step 5$: After acquiring the high-accuracy absolute phase, the 3D reconstruction can be carried out with the calibration parameters \cite{yin2012calibration} between the two cameras (see \textbf{Appendix B} for details).

\section{Experiments}
To verify the effectiveness of proposed approach, we construct a dual-camera system, which includes a LightCrafter 4500Pro ($912 \times 1140$ resolution) and two Basler acA640-750um cameras ($640 \times 480$ resolution). 48-period PS fringe patterns are used in our experiments. The size of the measuring field is about $240mm \times 200mm$.

To train our networks, we collect training datasets from 1001 different scenarios. With  training of hundreds of epochs, the training and validation loss of the networks converge without overfitting. We provide further details of collection of training data and the training process of the neural network in \textbf{Appendix C}.

\subsection{Qualitative evaluation}
To test the effectiveness of our approach, we firstly measure four static scenarios, containing single or multiple isolated objects with complex shapes, which are not in the training and verification datasets. We use four methods to measure these scenes. The first method is to use PS to obtain wrapped phase, and use triple-camera SPU and ADC to obtain absolute phase (the results obtained by which are taken as the ground-truth data); the second is to use PS to obtain wrapped phase, and use dual-camera SPU and conventional depth constraint strategy to obtain absolute phase; the third is to use PS to obtain wrapped phase, and directly use the reference phase to unwrap the phase; the fourth is our approach. The measurement results are shown in Fig. \ref{fig3}. It can be seen from the results of the second method that the conventional dual-camera SPU and depth constraints are insufficient to unwrap the phase of high-frequency fringes. The parts marked by the black dotted boxes in Fig. \ref{fig3} show the phase unwrapping errors of the third method, from which we can see that the reference plane can only unwrap the wrapped phase in a limited range, which is between $-\pi$ and $\pi$ of the absolute phase of the reference plane. While with our approach, the ambiguity of the wrapped phase can be accurately eliminated in a large depth range. In addition, our deep-learning-assisted approach can yield high-quality reconstruction results, almost the same quality as those obtained by conventional PS, triple-camera SPU and ADC methods.

We also test four continuously moving scenarios to demonstrate the superiority of our approach in dynamic target measurement (note that all our training and validation datasets are collected in static scenes). The measurement results are shown in Fig. \ref{fig4} (Multimedia view). It can be seen from the left three columns of Fig. \ref{fig4} (Multimedia view) that the multi-frame imaging characteristics of PS algorithm lead to obvious motion-induced artifacts in the reconstruction results when encountering moving objects. In addition, due to the sensitivity to phase errors, the results acquired by SPU obviously perform worse. Because of the single-shot nature of our approach, the measurement can be performed uninterruptedly without being affected by motion artifacts for dynamic scenarios, as shown in right most column of Fig. \ref{fig4} (Multimedia view).
\begin{figure}
	\centering
	\includegraphics[width=1\linewidth]{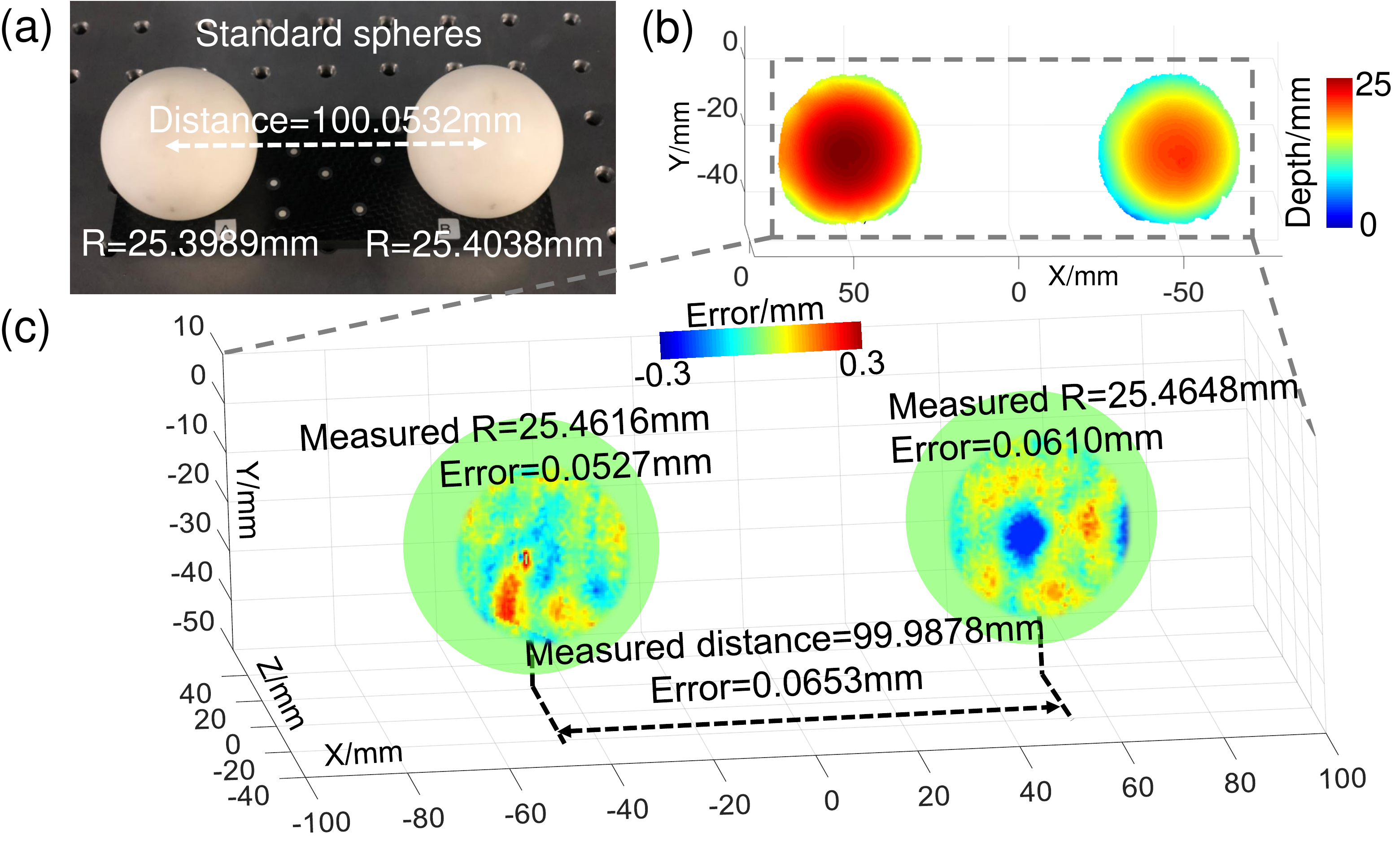}
	\caption{Quantitative analysis of our method. (a) The measured standard spheres. (b) 3D reconstruction result of our method. (c) The error distribution of the measured standard spheres.}
	\label{fig5}
\end{figure}

\subsection{Quantitative evaluation}
To quantitatively estimate the reconstruction accuracy of our approach, we measure two standard spheres, whose radii are 25.3989 $mm$ and 25.4038 $mm$ respectively, and the center-to-center distance is 100.0532 $mm$, with the uncertainty of 1.1 $\mu m$. Their form errors are 1.8 $\mu m$ and 3.5 $\mu m$ respectively. The measurement result is shown in Fig. \ref{fig5}(b). We perform sphere fitting to measured results of two spheres, and their errors are shown in Fig. \ref{fig5}(c). The radii of the reconstructed spheres are 25.4616 $mm$ and 25.4648 $mm$, with deviations of 52.7 $\mu m$ and 61.0 $\mu m$, respectively. The measured center distance is 99.9878 $mm$, with an error of 65.3 $\mu m$. This experiment validates that our method can provide high-quality 3D measurements with fewer cameras, fewer projection images, and simpler algorithms.

\section{Conclusions and discussions}
\subsection{Conclusions}
In this work, we presented a deep-learning-enabled geometric constraints and phase unwrapping approach for single-shot absolute 3D shape measurement. Our approach avoids the shortcomings of many traditional methods, such as the trade-off of efficiency and accuracy of conventional phase retrieval method, and the trade-off of SPU in phase unwrapping robustness, large measurement range and the use of high-frequency fringe patterns. On the premise of single-frame projection, our method can solve the phase ambiguity problem of dense fringe in a larger measurement range with less perspective information and simpler algorithms. We believe the proposed approach provides an important guidance for high-accuracy, motion-artifacts-free absolute 3D shape measurement for complicated object in high-speed scenarios.
\begin{figure}
	\centering
	\includegraphics[width=1\linewidth]{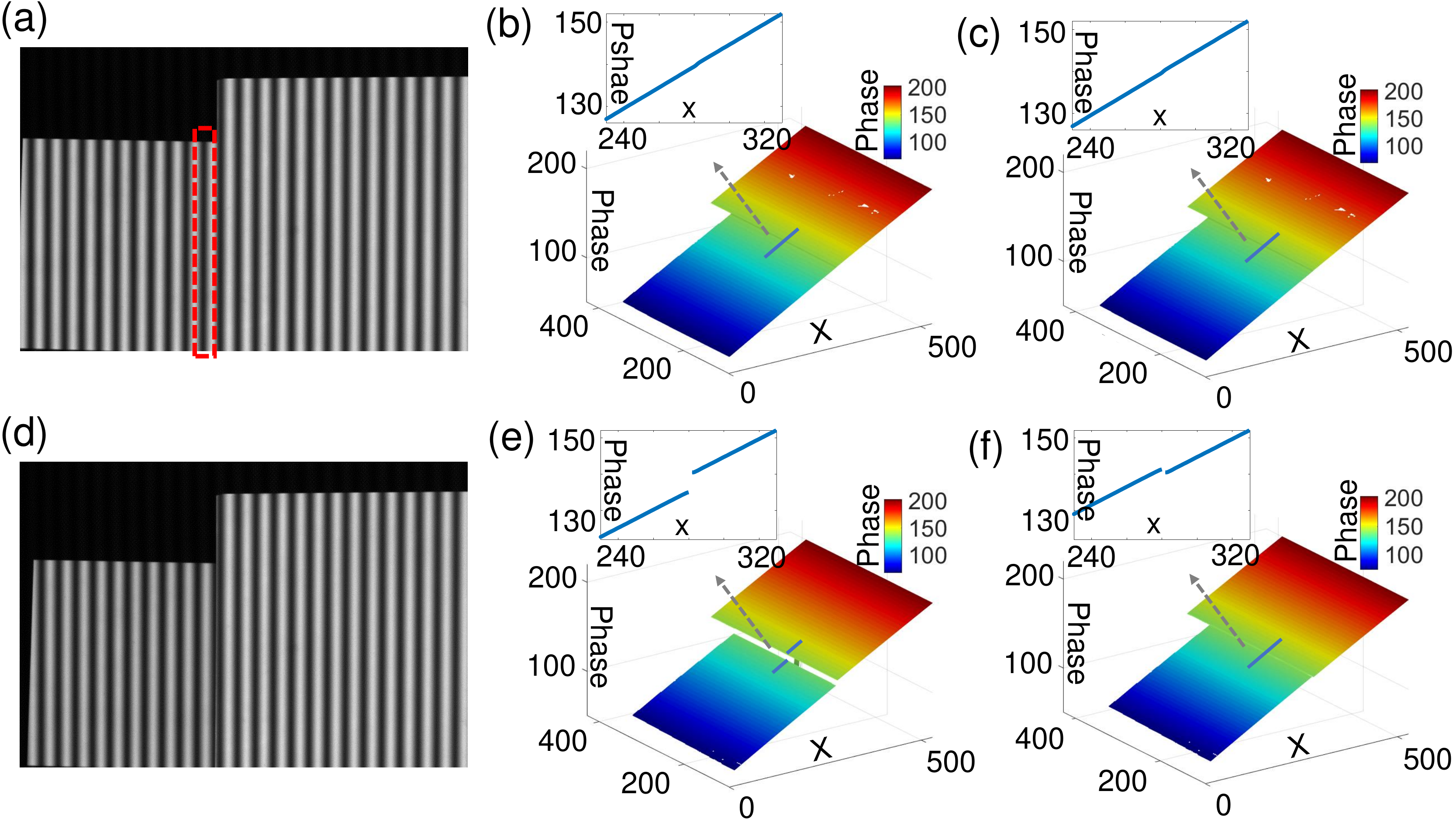}
	\caption{Analysis of the limitations of our method. (a) Image of two flat plates captured by Camera 1 (no ambiguity in the 2D image). (b) Absolute phase of two plates in (a). (c) The result of two plates in (a) obtained by our method. (d) Image of two flat plates captured by Camera 1 (the depth discontinuity of the objects results in missing order (the fringe in the red dotted box in (a)) and continuity artifact from the camera view). (e) Absolute phase of two plates in (d). (f) The result of two plates in (d) obtained by our method.}
	\label{fig6}
\end{figure}

\subsection{Discussions}
For traditional methods, one usually proceeds step by step based on prior knowledge. For example, for SPU, firstly find 3D candidates, then use depth constraints to remove unreliable candidate points, thirdly project to another perspective, and finally perform phase similarity check. Due to the step-by-step process, all information such as spatial information, temporal information, and so on, is not effectively utilized. The comprehensive utilization of all valid data requires strong and professional prior knowledge, which is very difficult to complete. However, deep learning can make it. Through data training and learning, these problems can be effectively integrated into a comprehensive framework. In our work, this framework is a very organic one, which incorporates phase acquisition, geometric constraints, and phase unwrapping. These methods in the framework are no longer reproduced step by step as traditionally, but are organically integrated together. However, since the data source of our method are 2D images, when the image itself is ambiguous, deep learning is by no means always reliable. For example, when the large depth discontinuity of the object results in missing order and continuity artifact from the camera view (Fig. \ref{fig6}), such inherent ambiguity in the captured fringe pattern cannot be resolved by deep learning techniques without additional auxiliary information such fringe images of different frequencies. In the future, we will further integrate the physical model into FPP based on deep learning, and construct FPP driven by data and physics.

\section*{Acknowledgements}
This study is supported by National Natural Science Foundation of China (61722506, 61705105, 11574152), National Key R$\&$D Program of China (2017YFF0106403), Outstanding Youth Foundation of Jiangsu Province (BK20170034), Fundamental Research Funds for the Central Universities (30917011204, 30919011222), and Leading Technology of Jiangsu Basic Research Plan (BK20192003).

\section*{Appendix}
\subsection*{Appendix A. Architecture of the neural networks}
We take CNN1 as example to reveal the internal structure of the constructed networks, as shown in the upper right part of Fig. \ref{fig2}. A 3D tensor with size $(H,W,C_0)$ is used as the input of the network, where $(H,W)$ is the size of the input images, and $C_0$ represents the number the input images. For each convolutional layer, the kernel size is $3 \times 3$ with convolution stride one, zero-padding is used to control the spatial size of the output, and the output is a 3D tensor of shape $(H,W,C)$, where  $C{\rm{ = }}64$ represents the number of filters used in each convolutional layer. In the first path of CNN1, the input is processed by a convolutional layer, followed by a group of residual blocks (containing four residual blocks) and another convolutional layer. Each residual block consists of 2 sets of convolutional layer activated by rectified linear unit (ReLU) stacked one above the other \cite{he2016deep} , which can solve the degradation of accuracy as the network becomes deeper and ease the training process. In the other three paths, the data is down-sampled by the pooling layers by two, four, and eight times, respectively, for better feature extraction, and then up-sampled by the upsampling blocks to match the original size. The outputs of four paths are concatenated into a tensor with quad channels. Finally, two channels are generated in the last convolution layer (one channel is generated in CNN2). Except for the last convolutional layer which is activated linearly, the rest ones use the ReLU as activation function. The mean-squared-errors of the outputs with respect to the ground truth are used as the loss function, and the adaptive moment estimation \cite{kingma2014adam} is utilized to tune the parameters for finding the minimum of the loss function.

\subsection*{Appendix B. System calibration and 3D reconstruction}
After acquiring the high-accuracy absolute phase, the matching points of two cameras can be uniquely identified. Then the 3D reconstruction can be carried out with the pre-calibration parameters between the two cameras. The reason why we utilize two cameras for reconstruction instead of one camera and one projector is that the multi-camera system can automatically cancel nonlinearity errors \cite{lohry2014absolute}. The calibration parameters, which contain the intrinsic, extrinsic, and distortion parameters of the cameras are calibrated based on MATLAB Calibration toolbox, and optimized with bundle adjustment \cite{yin2012calibration,huang2013camera}. 
\begin{figure}
	\centering
	\includegraphics[width=1\linewidth]{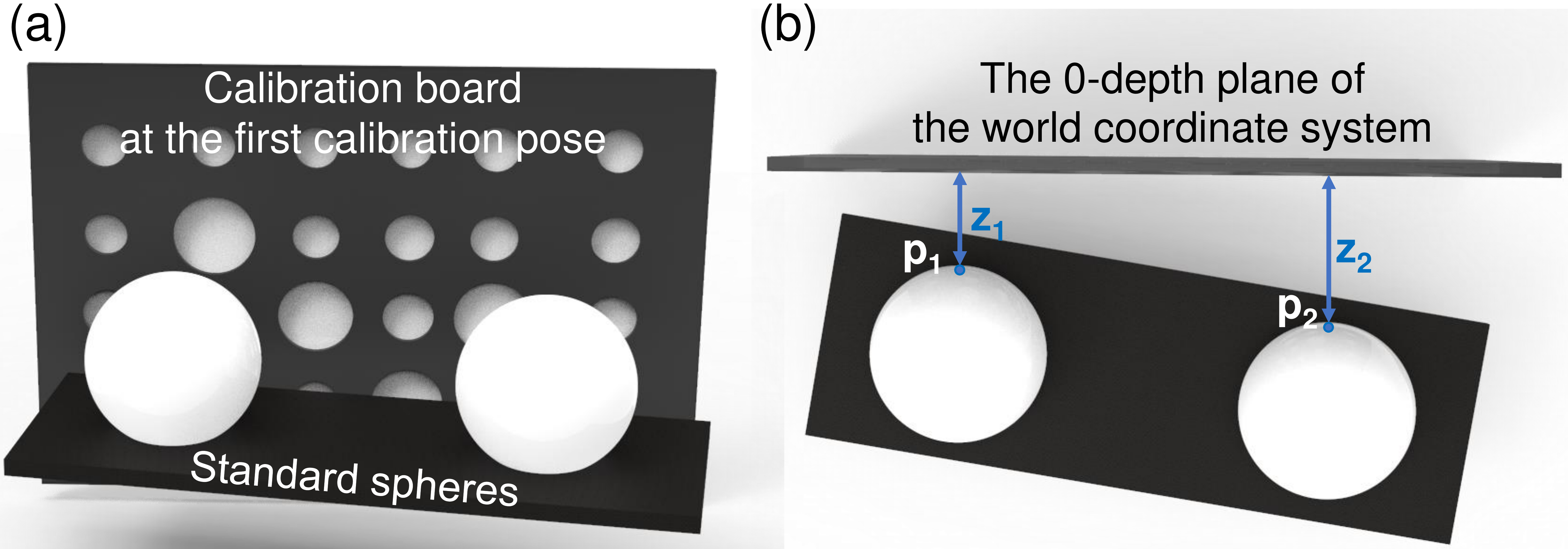}
	\caption{The relative position of the standard spheres and the calibration board at the first calibration pose.}
	\label{fig7}
\end{figure}

The reconstructed 3D coordinates are in the world coordinate system, the 0-depth plane of which corresponds to the position of the first calibration pose. For example, when the relationship between the position of a pair of standard spheres and the first calibration pose (the 0-depth plane of the world coordinate system) is as shown in Fig. \ref{fig7}, where Fig. \ref{fig7} (a) is the front view of the standard spheres and the calibration board in the first calibration pose and Fig. \ref{fig7}(b) is their top view, the depths of points $p_1$ and $p_2$ are $z_1$ and $z_2$, respectively.

\subsection*{Appendix C. Training the neural networks}
\begin{figure}
	\includegraphics[width=1\linewidth]{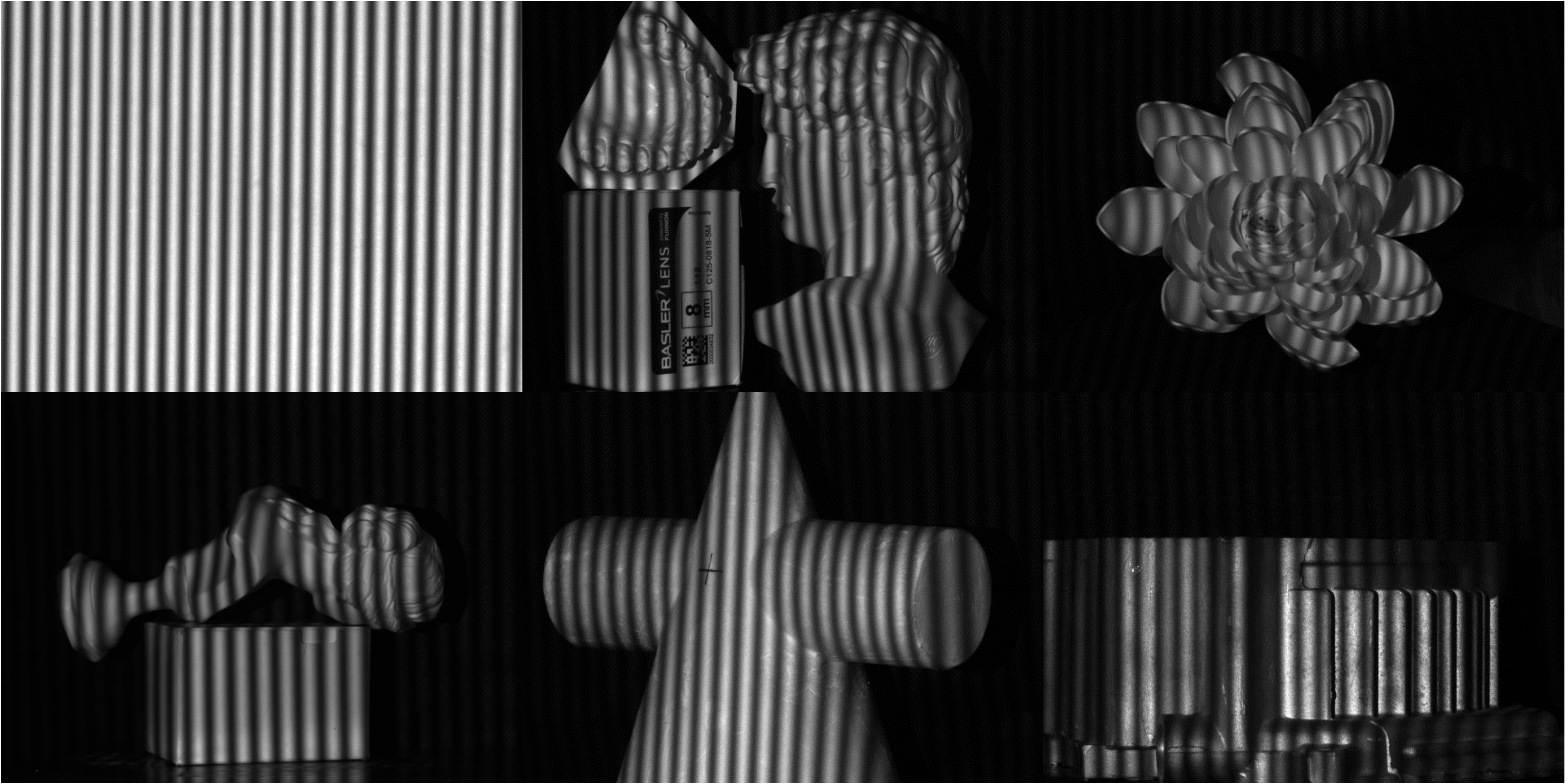}
	\caption{Six representative scenarios from total 1001 training datasets.}
	\label{fig8}
\end{figure}

To collect the training datasets, different types of simple and complex objects are arbitrarily combined and rotated 360 degrees to generate 1001 diverse scenes. Figure \ref{fig8} shows six representative scenarios from total 1001 training datasets, the first of which is the reference plane. Considering the following comparative experiments (verifying that our approach using only two perspectives can perform better than SPU using three cameras in dynamic scenes), we collect data from three views, each set of which consists of $3$-step PS fringe patterns captured by three cameras. Within each set of data, we calculate the ground-truth numerator $M$ and denominator $D$ by the 3-step PS algorithm, and obtain the fringe order maps by using triple-camera SPU and ADC (note that the fringe orders can also be acquired through only a single camera, by projecting multiple fringe patterns of different frequencies and using TPU). Before being fed into the networks, the fringe images are divided by 255 for normalization, and the fringe order maps are divided by the number of the used fringes (48) for normalization, which make the learning process easier for the network. When training the CNNs, 800 sets of data are used for training and 200 sets are used for verification. The training and verification datasets have been uploaded to the figshare (DOI:10.6084/m9.figshare.11926809; \href{https://figshare.com/s/f150a36191045e0c1bef}{https://figshare.com/s/f150a36191045e0c1bef}). 

\begin{figure}
	\centering
	\includegraphics[width=1\linewidth]{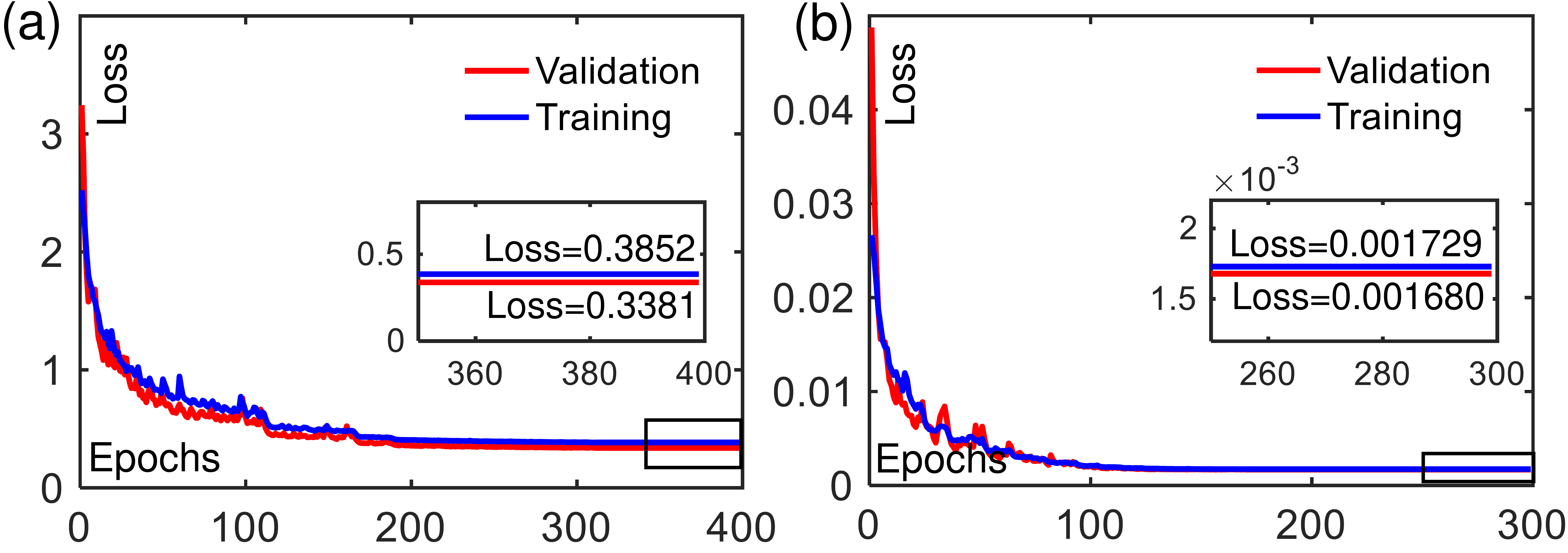}
	\caption{Loss curves of the training and validation set for (a) CNN1 and (b) CNN2.}
	\label{fig9}
\end{figure}

Figure \ref{fig9} shows the loss curve distributions of the CNNs. For CNN1, the loss curves converge after about 200 epochs, and the training of 400 epochs takes 25.56 hours; for CNN2, the loss curves converge after 120 epochs, the training of 300 epochs takes 19.25 hours. It is noted that the loss scales of the two networks are different because their outputs are not in the same scale: the numerator $M$ and denominator $D$ can reach hundreds while the fringe orders $k$ are normalized.

\nocite{*}
\bibliography{aipsamp}

\begin{thebibliography}{35}%
\makeatletter
\providecommand \@ifxundefined [1]{%
 \@ifx{#1\undefined}
}%
\providecommand \@ifnum [1]{%
 \ifnum #1\expandafter \@firstoftwo
 \else \expandafter \@secondoftwo
 \fi
}%
\providecommand \@ifx [1]{%
 \ifx #1\expandafter \@firstoftwo
 \else \expandafter \@secondoftwo
 \fi
}%
\providecommand \natexlab [1]{#1}%
\providecommand \enquote  [1]{``#1''}%
\providecommand \bibnamefont  [1]{#1}%
\providecommand \bibfnamefont [1]{#1}%
\providecommand \citenamefont [1]{#1}%
\providecommand \href@noop [0]{\@secondoftwo}%
\providecommand \href [0]{\begingroup \@sanitize@url \@href}%
\providecommand \@href[1]{\@@startlink{#1}\@@href}%
\providecommand \@@href[1]{\endgroup#1\@@endlink}%
\providecommand \@sanitize@url [0]{\catcode `\\12\catcode `\$12\catcode
  `\&12\catcode `\#12\catcode `\^12\catcode `\_12\catcode `\%12\relax}%
\providecommand \@@startlink[1]{}%
\providecommand \@@endlink[0]{}%
\providecommand \url  [0]{\begingroup\@sanitize@url \@url }%
\providecommand \@url [1]{\endgroup\@href {#1}{\urlprefix }}%
\providecommand \urlprefix  [0]{URL }%
\providecommand \Eprint [0]{\href }%
\providecommand \doibase [0]{http://dx.doi.org/}%
\providecommand \selectlanguage [0]{\@gobble}%
\providecommand \bibinfo  [0]{\@secondoftwo}%
\providecommand \bibfield  [0]{\@secondoftwo}%
\providecommand \translation [1]{[#1]}%
\providecommand \BibitemOpen [0]{}%
\providecommand \bibitemStop [0]{}%
\providecommand \bibitemNoStop [0]{.\EOS\space}%
\providecommand \EOS [0]{\spacefactor3000\relax}%
\providecommand \BibitemShut  [1]{\csname bibitem#1\endcsname}%
\let\auto@bib@innerbib\@empty
\bibitem [{\citenamefont {Salvi}\ \emph {et~al.}(2010)\citenamefont {Salvi},
  \citenamefont {Fernandez}, \citenamefont {Pribanic},\ and\ \citenamefont
  {Llado}}]{salvi2010state}%
  \BibitemOpen
  \bibfield  {author} {\bibinfo {author} {\bibfnamefont {J.}~\bibnamefont
  {Salvi}}, \bibinfo {author} {\bibfnamefont {S.}~\bibnamefont {Fernandez}},
  \bibinfo {author} {\bibfnamefont {T.}~\bibnamefont {Pribanic}}, \ and\
  \bibinfo {author} {\bibfnamefont {X.}~\bibnamefont {Llado}},\ }\bibfield
  {title} {\enquote {\bibinfo {title} {A state of the art in structured light
  patterns for surface profilometry},}\ }\href@noop {} {\bibfield  {journal}
  {\bibinfo  {journal} {Pattern recognition}\ }\textbf {\bibinfo {volume}
  {43}},\ \bibinfo {pages} {2666--2680} (\bibinfo {year} {2010})}\BibitemShut
  {NoStop}%
\bibitem [{\citenamefont {Gorthi}\ and\ \citenamefont
  {Rastogi}(2010)}]{gorthi2010fringe}%
  \BibitemOpen
  \bibfield  {author} {\bibinfo {author} {\bibfnamefont {S.~S.}\ \bibnamefont
  {Gorthi}}\ and\ \bibinfo {author} {\bibfnamefont {P.}~\bibnamefont
  {Rastogi}},\ }\bibfield  {title} {\enquote {\bibinfo {title} {Fringe
  projection techniques: whither we are?}}\ }\href@noop {} {\bibfield
  {journal} {\bibinfo  {journal} {Optics and lasers in engineering}\ }\textbf
  {\bibinfo {volume} {48}},\ \bibinfo {pages} {133--140} (\bibinfo {year}
  {2010})}\BibitemShut {NoStop}%
\bibitem [{\citenamefont {Zhang}, \citenamefont {Van Der~Weide},\ and\
  \citenamefont {Oliver}(2010)}]{zhang2010superfast}%
  \BibitemOpen
  \bibfield  {author} {\bibinfo {author} {\bibfnamefont {S.}~\bibnamefont
  {Zhang}}, \bibinfo {author} {\bibfnamefont {D.}~\bibnamefont {Van
  Der~Weide}}, \ and\ \bibinfo {author} {\bibfnamefont {J.}~\bibnamefont
  {Oliver}},\ }\bibfield  {title} {\enquote {\bibinfo {title} {Superfast
  phase-shifting method for 3-d shape measurement},}\ }\href@noop {} {\bibfield
   {journal} {\bibinfo  {journal} {Optics express}\ }\textbf {\bibinfo {volume}
  {18}},\ \bibinfo {pages} {9684--9689} (\bibinfo {year} {2010})}\BibitemShut
  {NoStop}%
\bibitem [{\citenamefont {Zuo}\ \emph {et~al.}(2013)\citenamefont {Zuo},
  \citenamefont {Chen}, \citenamefont {Gu}, \citenamefont {Feng}, \citenamefont
  {Feng}, \citenamefont {Li},\ and\ \citenamefont {Shen}}]{zuo2013high}%
  \BibitemOpen
  \bibfield  {author} {\bibinfo {author} {\bibfnamefont {C.}~\bibnamefont
  {Zuo}}, \bibinfo {author} {\bibfnamefont {Q.}~\bibnamefont {Chen}}, \bibinfo
  {author} {\bibfnamefont {G.}~\bibnamefont {Gu}}, \bibinfo {author}
  {\bibfnamefont {S.}~\bibnamefont {Feng}}, \bibinfo {author} {\bibfnamefont
  {F.}~\bibnamefont {Feng}}, \bibinfo {author} {\bibfnamefont {R.}~\bibnamefont
  {Li}}, \ and\ \bibinfo {author} {\bibfnamefont {G.}~\bibnamefont {Shen}},\
  }\bibfield  {title} {\enquote {\bibinfo {title} {High-speed three-dimensional
  shape measurement for dynamic scenes using bi-frequency tripolar
  pulse-width-modulation fringe projection},}\ }\href@noop {} {\bibfield
  {journal} {\bibinfo  {journal} {Optics and Lasers in Engineering}\ }\textbf
  {\bibinfo {volume} {51}},\ \bibinfo {pages} {953--960} (\bibinfo {year}
  {2013})}\BibitemShut {NoStop}%
\bibitem [{\citenamefont {Tao}\ \emph {et~al.}(2018)\citenamefont {Tao},
  \citenamefont {Chen}, \citenamefont {Feng}, \citenamefont {Qian},
  \citenamefont {Hu}, \citenamefont {Huang},\ and\ \citenamefont
  {Zuo}}]{tao2018high}%
  \BibitemOpen
  \bibfield  {author} {\bibinfo {author} {\bibfnamefont {T.}~\bibnamefont
  {Tao}}, \bibinfo {author} {\bibfnamefont {Q.}~\bibnamefont {Chen}}, \bibinfo
  {author} {\bibfnamefont {S.}~\bibnamefont {Feng}}, \bibinfo {author}
  {\bibfnamefont {J.}~\bibnamefont {Qian}}, \bibinfo {author} {\bibfnamefont
  {Y.}~\bibnamefont {Hu}}, \bibinfo {author} {\bibfnamefont {L.}~\bibnamefont
  {Huang}}, \ and\ \bibinfo {author} {\bibfnamefont {C.}~\bibnamefont {Zuo}},\
  }\bibfield  {title} {\enquote {\bibinfo {title} {High-speed real-time 3d
  shape measurement based on adaptive depth constraint},}\ }\href@noop {}
  {\bibfield  {journal} {\bibinfo  {journal} {Optics express}\ }\textbf
  {\bibinfo {volume} {26}},\ \bibinfo {pages} {22440--22456} (\bibinfo {year}
  {2018})}\BibitemShut {NoStop}%
\bibitem [{\citenamefont {Feng}\ \emph
  {et~al.}(2018{\natexlab{a}})\citenamefont {Feng}, \citenamefont {Zuo},
  \citenamefont {Tao}, \citenamefont {Hu}, \citenamefont {Zhang}, \citenamefont
  {Chen},\ and\ \citenamefont {Gu}}]{feng2018robust}%
  \BibitemOpen
  \bibfield  {author} {\bibinfo {author} {\bibfnamefont {S.}~\bibnamefont
  {Feng}}, \bibinfo {author} {\bibfnamefont {C.}~\bibnamefont {Zuo}}, \bibinfo
  {author} {\bibfnamefont {T.}~\bibnamefont {Tao}}, \bibinfo {author}
  {\bibfnamefont {Y.}~\bibnamefont {Hu}}, \bibinfo {author} {\bibfnamefont
  {M.}~\bibnamefont {Zhang}}, \bibinfo {author} {\bibfnamefont
  {Q.}~\bibnamefont {Chen}}, \ and\ \bibinfo {author} {\bibfnamefont
  {G.}~\bibnamefont {Gu}},\ }\bibfield  {title} {\enquote {\bibinfo {title}
  {Robust dynamic 3-d measurements with motion-compensated phase-shifting
  profilometry},}\ }\href@noop {} {\bibfield  {journal} {\bibinfo  {journal}
  {Optics and Lasers in Engineering}\ }\textbf {\bibinfo {volume} {103}},\
  \bibinfo {pages} {127--138} (\bibinfo {year}
  {2018}{\natexlab{a}})}\BibitemShut {NoStop}%
\bibitem [{\citenamefont {Feng}\ \emph
  {et~al.}(2018{\natexlab{b}})\citenamefont {Feng}, \citenamefont {Zhang},
  \citenamefont {Zuo}, \citenamefont {Tao}, \citenamefont {Chen},\ and\
  \citenamefont {Gu}}]{feng2018high}%
  \BibitemOpen
  \bibfield  {author} {\bibinfo {author} {\bibfnamefont {S.}~\bibnamefont
  {Feng}}, \bibinfo {author} {\bibfnamefont {L.}~\bibnamefont {Zhang}},
  \bibinfo {author} {\bibfnamefont {C.}~\bibnamefont {Zuo}}, \bibinfo {author}
  {\bibfnamefont {T.}~\bibnamefont {Tao}}, \bibinfo {author} {\bibfnamefont
  {Q.}~\bibnamefont {Chen}}, \ and\ \bibinfo {author} {\bibfnamefont
  {G.}~\bibnamefont {Gu}},\ }\bibfield  {title} {\enquote {\bibinfo {title}
  {High dynamic range 3d measurements with fringe projection profilometry: a
  review},}\ }\href@noop {} {\bibfield  {journal} {\bibinfo  {journal}
  {Measurement Science and Technology}\ }\textbf {\bibinfo {volume} {29}},\
  \bibinfo {pages} {122001} (\bibinfo {year} {2018}{\natexlab{b}})}\BibitemShut
  {NoStop}%
\bibitem [{\citenamefont {Zuo}\ \emph {et~al.}(2018{\natexlab{a}})\citenamefont
  {Zuo}, \citenamefont {Tao}, \citenamefont {Feng}, \citenamefont {Huang},
  \citenamefont {Asundi},\ and\ \citenamefont {Chen}}]{zuo2018micro}%
  \BibitemOpen
  \bibfield  {author} {\bibinfo {author} {\bibfnamefont {C.}~\bibnamefont
  {Zuo}}, \bibinfo {author} {\bibfnamefont {T.}~\bibnamefont {Tao}}, \bibinfo
  {author} {\bibfnamefont {S.}~\bibnamefont {Feng}}, \bibinfo {author}
  {\bibfnamefont {L.}~\bibnamefont {Huang}}, \bibinfo {author} {\bibfnamefont
  {A.}~\bibnamefont {Asundi}}, \ and\ \bibinfo {author} {\bibfnamefont
  {Q.}~\bibnamefont {Chen}},\ }\bibfield  {title} {\enquote {\bibinfo {title}
  {Micro fourier transform profilometry ($\mu$ftp): 3d shape measurement at
  10,000 frames per second},}\ }\href@noop {} {\bibfield  {journal} {\bibinfo
  {journal} {Optics and Lasers in Engineering}\ }\textbf {\bibinfo {volume}
  {102}},\ \bibinfo {pages} {70--91} (\bibinfo {year}
  {2018}{\natexlab{a}})}\BibitemShut {NoStop}%
\bibitem [{\citenamefont {Qian}\ \emph
  {et~al.}(2019{\natexlab{a}})\citenamefont {Qian}, \citenamefont {Feng},
  \citenamefont {Tao}, \citenamefont {Hu}, \citenamefont {Liu}, \citenamefont
  {Wu}, \citenamefont {Chen},\ and\ \citenamefont {Zuo}}]{qian2019high}%
  \BibitemOpen
  \bibfield  {author} {\bibinfo {author} {\bibfnamefont {J.}~\bibnamefont
  {Qian}}, \bibinfo {author} {\bibfnamefont {S.}~\bibnamefont {Feng}}, \bibinfo
  {author} {\bibfnamefont {T.}~\bibnamefont {Tao}}, \bibinfo {author}
  {\bibfnamefont {Y.}~\bibnamefont {Hu}}, \bibinfo {author} {\bibfnamefont
  {K.}~\bibnamefont {Liu}}, \bibinfo {author} {\bibfnamefont {S.}~\bibnamefont
  {Wu}}, \bibinfo {author} {\bibfnamefont {Q.}~\bibnamefont {Chen}}, \ and\
  \bibinfo {author} {\bibfnamefont {C.}~\bibnamefont {Zuo}},\ }\bibfield
  {title} {\enquote {\bibinfo {title} {High-resolution real-time 360$^\circ $
  3d model reconstruction of a handheld object with fringe projection
  profilometry},}\ }\href@noop {} {\bibfield  {journal} {\bibinfo  {journal}
  {Optics letters}\ }\textbf {\bibinfo {volume} {44}},\ \bibinfo {pages}
  {5751--5754} (\bibinfo {year} {2019}{\natexlab{a}})}\BibitemShut {NoStop}%
\bibitem [{\citenamefont {Feng}\ \emph
  {et~al.}(2019{\natexlab{a}})\citenamefont {Feng}, \citenamefont {Chen},
  \citenamefont {Gu}, \citenamefont {Tao}, \citenamefont {Zhang}, \citenamefont
  {Hu}, \citenamefont {Yin},\ and\ \citenamefont {Zuo}}]{feng2019fringe}%
  \BibitemOpen
  \bibfield  {author} {\bibinfo {author} {\bibfnamefont {S.}~\bibnamefont
  {Feng}}, \bibinfo {author} {\bibfnamefont {Q.}~\bibnamefont {Chen}}, \bibinfo
  {author} {\bibfnamefont {G.}~\bibnamefont {Gu}}, \bibinfo {author}
  {\bibfnamefont {T.}~\bibnamefont {Tao}}, \bibinfo {author} {\bibfnamefont
  {L.}~\bibnamefont {Zhang}}, \bibinfo {author} {\bibfnamefont
  {Y.}~\bibnamefont {Hu}}, \bibinfo {author} {\bibfnamefont {W.}~\bibnamefont
  {Yin}}, \ and\ \bibinfo {author} {\bibfnamefont {C.}~\bibnamefont {Zuo}},\
  }\bibfield  {title} {\enquote {\bibinfo {title} {Fringe pattern analysis
  using deep learning},}\ }\href@noop {} {\bibfield  {journal} {\bibinfo
  {journal} {Advanced Photonics}\ }\textbf {\bibinfo {volume} {1}},\ \bibinfo
  {pages} {025001} (\bibinfo {year} {2019}{\natexlab{a}})}\BibitemShut
  {NoStop}%
\bibitem [{\citenamefont {Feng}\ \emph
  {et~al.}(2019{\natexlab{b}})\citenamefont {Feng}, \citenamefont {Zuo},
  \citenamefont {Yin}, \citenamefont {Gu},\ and\ \citenamefont
  {Chen}}]{feng2019micro}%
  \BibitemOpen
  \bibfield  {author} {\bibinfo {author} {\bibfnamefont {S.}~\bibnamefont
  {Feng}}, \bibinfo {author} {\bibfnamefont {C.}~\bibnamefont {Zuo}}, \bibinfo
  {author} {\bibfnamefont {W.}~\bibnamefont {Yin}}, \bibinfo {author}
  {\bibfnamefont {G.}~\bibnamefont {Gu}}, \ and\ \bibinfo {author}
  {\bibfnamefont {Q.}~\bibnamefont {Chen}},\ }\bibfield  {title} {\enquote
  {\bibinfo {title} {Micro deep learning profilometry for high-speed 3d surface
  imaging},}\ }\href@noop {} {\bibfield  {journal} {\bibinfo  {journal} {Optics
  and Lasers in Engineering}\ }\textbf {\bibinfo {volume} {121}},\ \bibinfo
  {pages} {416--427} (\bibinfo {year} {2019}{\natexlab{b}})}\BibitemShut
  {NoStop}%
\bibitem [{\citenamefont {Zuo}\ \emph {et~al.}(2016)\citenamefont {Zuo},
  \citenamefont {Huang}, \citenamefont {Zhang}, \citenamefont {Chen},\ and\
  \citenamefont {Asundi}}]{zuo2016temporal}%
  \BibitemOpen
  \bibfield  {author} {\bibinfo {author} {\bibfnamefont {C.}~\bibnamefont
  {Zuo}}, \bibinfo {author} {\bibfnamefont {L.}~\bibnamefont {Huang}}, \bibinfo
  {author} {\bibfnamefont {M.}~\bibnamefont {Zhang}}, \bibinfo {author}
  {\bibfnamefont {Q.}~\bibnamefont {Chen}}, \ and\ \bibinfo {author}
  {\bibfnamefont {A.}~\bibnamefont {Asundi}},\ }\bibfield  {title} {\enquote
  {\bibinfo {title} {Temporal phase unwrapping algorithms for fringe projection
  profilometry: A comparative review},}\ }\href@noop {} {\bibfield  {journal}
  {\bibinfo  {journal} {Optics and Lasers in Engineering}\ }\textbf {\bibinfo
  {volume} {85}},\ \bibinfo {pages} {84--103} (\bibinfo {year}
  {2016})}\BibitemShut {NoStop}%
\bibitem [{\citenamefont {Zhang}, \citenamefont {Towers},\ and\ \citenamefont
  {Towers}(2006)}]{zhang2006time}%
  \BibitemOpen
  \bibfield  {author} {\bibinfo {author} {\bibfnamefont {Z.}~\bibnamefont
  {Zhang}}, \bibinfo {author} {\bibfnamefont {C.~E.}\ \bibnamefont {Towers}}, \
  and\ \bibinfo {author} {\bibfnamefont {D.~P.}\ \bibnamefont {Towers}},\
  }\bibfield  {title} {\enquote {\bibinfo {title} {Time efficient color fringe
  projection system for 3d shape and color using optimum 3-frequency
  selection},}\ }\href@noop {} {\bibfield  {journal} {\bibinfo  {journal}
  {Optics express}\ }\textbf {\bibinfo {volume} {14}},\ \bibinfo {pages}
  {6444--6455} (\bibinfo {year} {2006})}\BibitemShut {NoStop}%
\bibitem [{\citenamefont {Weise}, \citenamefont {Leibe},\ and\ \citenamefont
  {Van~Gool}(2007)}]{weise2007fast}%
  \BibitemOpen
  \bibfield  {author} {\bibinfo {author} {\bibfnamefont {T.}~\bibnamefont
  {Weise}}, \bibinfo {author} {\bibfnamefont {B.}~\bibnamefont {Leibe}}, \ and\
  \bibinfo {author} {\bibfnamefont {L.}~\bibnamefont {Van~Gool}},\ }\bibfield
  {title} {\enquote {\bibinfo {title} {Fast 3d scanning with automatic motion
  compensation},}\ }in\ \href@noop {} {\emph {\bibinfo {booktitle} {2007 IEEE
  Conference on Computer Vision and Pattern Recognition}}}\ (\bibinfo
  {organization} {IEEE},\ \bibinfo {year} {2007})\ pp.\ \bibinfo {pages}
  {1--8}\BibitemShut {NoStop}%
\bibitem [{\citenamefont {Su}\ and\ \citenamefont
  {Chen}(2004)}]{su2004reliability}%
  \BibitemOpen
  \bibfield  {author} {\bibinfo {author} {\bibfnamefont {X.}~\bibnamefont
  {Su}}\ and\ \bibinfo {author} {\bibfnamefont {W.}~\bibnamefont {Chen}},\
  }\bibfield  {title} {\enquote {\bibinfo {title} {Reliability-guided phase
  unwrapping algorithm: a review},}\ }\href@noop {} {\bibfield  {journal}
  {\bibinfo  {journal} {Optics and Lasers in Engineering}\ }\textbf {\bibinfo
  {volume} {42}},\ \bibinfo {pages} {245--261} (\bibinfo {year}
  {2004})}\BibitemShut {NoStop}%
\bibitem [{\citenamefont {Garcia}\ and\ \citenamefont
  {Zakhor}(2012)}]{garcia2012consistent}%
  \BibitemOpen
  \bibfield  {author} {\bibinfo {author} {\bibfnamefont {R.~R.}\ \bibnamefont
  {Garcia}}\ and\ \bibinfo {author} {\bibfnamefont {A.}~\bibnamefont
  {Zakhor}},\ }\bibfield  {title} {\enquote {\bibinfo {title} {Consistent
  stereo-assisted absolute phase unwrapping methods for structured light
  systems},}\ }\href@noop {} {\bibfield  {journal} {\bibinfo  {journal} {IEEE
  Journal of selected topics in Signal Processing}\ }\textbf {\bibinfo {volume}
  {6}},\ \bibinfo {pages} {411--424} (\bibinfo {year} {2012})}\BibitemShut
  {NoStop}%
\bibitem [{\citenamefont {Lohry}\ and\ \citenamefont
  {Zhang}(2014)}]{lohry2014high}%
  \BibitemOpen
  \bibfield  {author} {\bibinfo {author} {\bibfnamefont {W.}~\bibnamefont
  {Lohry}}\ and\ \bibinfo {author} {\bibfnamefont {S.}~\bibnamefont {Zhang}},\
  }\bibfield  {title} {\enquote {\bibinfo {title} {High-speed absolute
  three-dimensional shape measurement using three binary dithered patterns},}\
  }\href@noop {} {\bibfield  {journal} {\bibinfo  {journal} {Optics express}\
  }\textbf {\bibinfo {volume} {22}},\ \bibinfo {pages} {26752--26762} (\bibinfo
  {year} {2014})}\BibitemShut {NoStop}%
\bibitem [{\citenamefont {Tao}\ \emph {et~al.}(2016)\citenamefont {Tao},
  \citenamefont {Chen}, \citenamefont {Da}, \citenamefont {Feng}, \citenamefont
  {Hu},\ and\ \citenamefont {Zuo}}]{tao2016real}%
  \BibitemOpen
  \bibfield  {author} {\bibinfo {author} {\bibfnamefont {T.}~\bibnamefont
  {Tao}}, \bibinfo {author} {\bibfnamefont {Q.}~\bibnamefont {Chen}}, \bibinfo
  {author} {\bibfnamefont {J.}~\bibnamefont {Da}}, \bibinfo {author}
  {\bibfnamefont {S.}~\bibnamefont {Feng}}, \bibinfo {author} {\bibfnamefont
  {Y.}~\bibnamefont {Hu}}, \ and\ \bibinfo {author} {\bibfnamefont
  {C.}~\bibnamefont {Zuo}},\ }\bibfield  {title} {\enquote {\bibinfo {title}
  {Real-time 3-d shape measurement with composite phase-shifting fringes and
  multi-view system},}\ }\href@noop {} {\bibfield  {journal} {\bibinfo
  {journal} {Optics express}\ }\textbf {\bibinfo {volume} {24}},\ \bibinfo
  {pages} {20253--20269} (\bibinfo {year} {2016})}\BibitemShut {NoStop}%
\bibitem [{\citenamefont {Tao}\ \emph {et~al.}(2017)\citenamefont {Tao},
  \citenamefont {Chen}, \citenamefont {Feng}, \citenamefont {Hu}, \citenamefont
  {Zhang},\ and\ \citenamefont {Zuo}}]{tao2017high}%
  \BibitemOpen
  \bibfield  {author} {\bibinfo {author} {\bibfnamefont {T.}~\bibnamefont
  {Tao}}, \bibinfo {author} {\bibfnamefont {Q.}~\bibnamefont {Chen}}, \bibinfo
  {author} {\bibfnamefont {S.}~\bibnamefont {Feng}}, \bibinfo {author}
  {\bibfnamefont {Y.}~\bibnamefont {Hu}}, \bibinfo {author} {\bibfnamefont
  {M.}~\bibnamefont {Zhang}}, \ and\ \bibinfo {author} {\bibfnamefont
  {C.}~\bibnamefont {Zuo}},\ }\bibfield  {title} {\enquote {\bibinfo {title}
  {High-precision real-time 3d shape measurement based on a quad-camera
  system},}\ }\href@noop {} {\bibfield  {journal} {\bibinfo  {journal} {Journal
  of Optics}\ }\textbf {\bibinfo {volume} {20}},\ \bibinfo {pages} {014009}
  (\bibinfo {year} {2017})}\BibitemShut {NoStop}%
\bibitem [{\citenamefont {Br{\"a}uer-Burchardt}\ \emph
  {et~al.}(2011)\citenamefont {Br{\"a}uer-Burchardt}, \citenamefont {Munkelt},
  \citenamefont {Heinze}, \citenamefont {K{\"u}hmstedt},\ and\ \citenamefont
  {Notni}}]{brauer2011using}%
  \BibitemOpen
  \bibfield  {author} {\bibinfo {author} {\bibfnamefont {C.}~\bibnamefont
  {Br{\"a}uer-Burchardt}}, \bibinfo {author} {\bibfnamefont {C.}~\bibnamefont
  {Munkelt}}, \bibinfo {author} {\bibfnamefont {M.}~\bibnamefont {Heinze}},
  \bibinfo {author} {\bibfnamefont {P.}~\bibnamefont {K{\"u}hmstedt}}, \ and\
  \bibinfo {author} {\bibfnamefont {G.}~\bibnamefont {Notni}},\ }\bibfield
  {title} {\enquote {\bibinfo {title} {Using geometric constraints to solve the
  point correspondence problem in fringe projection based 3d measuring
  systems},}\ }in\ \href@noop {} {\emph {\bibinfo {booktitle} {International
  Conference on Image Analysis and Processing}}}\ (\bibinfo {organization}
  {Springer},\ \bibinfo {year} {2011})\ pp.\ \bibinfo {pages}
  {265--274}\BibitemShut {NoStop}%
\bibitem [{\citenamefont {Li}\ \emph {et~al.}(2013)\citenamefont {Li},
  \citenamefont {Zhong}, \citenamefont {Li}, \citenamefont {Zhou},\ and\
  \citenamefont {Shi}}]{li2013multiview}%
  \BibitemOpen
  \bibfield  {author} {\bibinfo {author} {\bibfnamefont {Z.}~\bibnamefont
  {Li}}, \bibinfo {author} {\bibfnamefont {K.}~\bibnamefont {Zhong}}, \bibinfo
  {author} {\bibfnamefont {Y.~F.}\ \bibnamefont {Li}}, \bibinfo {author}
  {\bibfnamefont {X.}~\bibnamefont {Zhou}}, \ and\ \bibinfo {author}
  {\bibfnamefont {Y.}~\bibnamefont {Shi}},\ }\bibfield  {title} {\enquote
  {\bibinfo {title} {Multiview phase shifting: a full-resolution and high-speed
  3d measurement framework for arbitrary shape dynamic objects},}\ }\href@noop
  {} {\bibfield  {journal} {\bibinfo  {journal} {Optics letters}\ }\textbf
  {\bibinfo {volume} {38}},\ \bibinfo {pages} {1389--1391} (\bibinfo {year}
  {2013})}\BibitemShut {NoStop}%
\bibitem [{\citenamefont {Liu}\ and\ \citenamefont
  {Kofman}(2017)}]{liu2017high}%
  \BibitemOpen
  \bibfield  {author} {\bibinfo {author} {\bibfnamefont {X.}~\bibnamefont
  {Liu}}\ and\ \bibinfo {author} {\bibfnamefont {J.}~\bibnamefont {Kofman}},\
  }\bibfield  {title} {\enquote {\bibinfo {title} {High-frequency background
  modulation fringe patterns based on a fringe-wavelength geometry-constraint
  model for 3d surface-shape measurement},}\ }\href@noop {} {\bibfield
  {journal} {\bibinfo  {journal} {Optics Express}\ }\textbf {\bibinfo {volume}
  {25}},\ \bibinfo {pages} {16618--16628} (\bibinfo {year} {2017})}\BibitemShut
  {NoStop}%
\bibitem [{\citenamefont {Qian}\ \emph
  {et~al.}(2019{\natexlab{b}})\citenamefont {Qian}, \citenamefont {Tao},
  \citenamefont {Feng}, \citenamefont {Chen},\ and\ \citenamefont
  {Zuo}}]{qian2019motion}%
  \BibitemOpen
  \bibfield  {author} {\bibinfo {author} {\bibfnamefont {J.}~\bibnamefont
  {Qian}}, \bibinfo {author} {\bibfnamefont {T.}~\bibnamefont {Tao}}, \bibinfo
  {author} {\bibfnamefont {S.}~\bibnamefont {Feng}}, \bibinfo {author}
  {\bibfnamefont {Q.}~\bibnamefont {Chen}}, \ and\ \bibinfo {author}
  {\bibfnamefont {C.}~\bibnamefont {Zuo}},\ }\bibfield  {title} {\enquote
  {\bibinfo {title} {Motion-artifact-free dynamic 3d shape measurement with
  hybrid fourier-transform phase-shifting profilometry},}\ }\href@noop {}
  {\bibfield  {journal} {\bibinfo  {journal} {Optics express}\ }\textbf
  {\bibinfo {volume} {27}},\ \bibinfo {pages} {2713--2731} (\bibinfo {year}
  {2019}{\natexlab{b}})}\BibitemShut {NoStop}%
\bibitem [{\citenamefont {Zuo}\ \emph {et~al.}(2018{\natexlab{b}})\citenamefont
  {Zuo}, \citenamefont {Feng}, \citenamefont {Huang}, \citenamefont {Tao},
  \citenamefont {Yin},\ and\ \citenamefont {Chen}}]{zuo2018phase}%
  \BibitemOpen
  \bibfield  {author} {\bibinfo {author} {\bibfnamefont {C.}~\bibnamefont
  {Zuo}}, \bibinfo {author} {\bibfnamefont {S.}~\bibnamefont {Feng}}, \bibinfo
  {author} {\bibfnamefont {L.}~\bibnamefont {Huang}}, \bibinfo {author}
  {\bibfnamefont {T.}~\bibnamefont {Tao}}, \bibinfo {author} {\bibfnamefont
  {W.}~\bibnamefont {Yin}}, \ and\ \bibinfo {author} {\bibfnamefont
  {Q.}~\bibnamefont {Chen}},\ }\bibfield  {title} {\enquote {\bibinfo {title}
  {Phase shifting algorithms for fringe projection profilometry: A review},}\
  }\href@noop {} {\bibfield  {journal} {\bibinfo  {journal} {Optics and Lasers
  in Engineering}\ }\textbf {\bibinfo {volume} {109}},\ \bibinfo {pages}
  {23--59} (\bibinfo {year} {2018}{\natexlab{b}})}\BibitemShut {NoStop}%
\bibitem [{\citenamefont {Su}\ and\ \citenamefont
  {Zhang}(2010)}]{su2010dynamic}%
  \BibitemOpen
  \bibfield  {author} {\bibinfo {author} {\bibfnamefont {X.}~\bibnamefont
  {Su}}\ and\ \bibinfo {author} {\bibfnamefont {Q.}~\bibnamefont {Zhang}},\
  }\bibfield  {title} {\enquote {\bibinfo {title} {Dynamic 3-d shape
  measurement method: a review},}\ }\href@noop {} {\bibfield  {journal}
  {\bibinfo  {journal} {Optics and Lasers in Engineering}\ }\textbf {\bibinfo
  {volume} {48}},\ \bibinfo {pages} {191--204} (\bibinfo {year}
  {2010})}\BibitemShut {NoStop}%
\bibitem [{\citenamefont {Huang}\ \emph {et~al.}(2010)\citenamefont {Huang},
  \citenamefont {Kemao}, \citenamefont {Pan},\ and\ \citenamefont
  {Asundi}}]{huang2010comparison}%
  \BibitemOpen
  \bibfield  {author} {\bibinfo {author} {\bibfnamefont {L.}~\bibnamefont
  {Huang}}, \bibinfo {author} {\bibfnamefont {Q.}~\bibnamefont {Kemao}},
  \bibinfo {author} {\bibfnamefont {B.}~\bibnamefont {Pan}}, \ and\ \bibinfo
  {author} {\bibfnamefont {A.~K.}\ \bibnamefont {Asundi}},\ }\bibfield  {title}
  {\enquote {\bibinfo {title} {Comparison of fourier transform, windowed
  fourier transform, and wavelet transform methods for phase extraction from a
  single fringe pattern in fringe projection profilometry},}\ }\href@noop {}
  {\bibfield  {journal} {\bibinfo  {journal} {Optics and Lasers in
  Engineering}\ }\textbf {\bibinfo {volume} {48}},\ \bibinfo {pages} {141--148}
  (\bibinfo {year} {2010})}\BibitemShut {NoStop}%
\bibitem [{\citenamefont {Van~der Jeught}\ and\ \citenamefont
  {Dirckx}(2019)}]{van2019deep}%
  \BibitemOpen
  \bibfield  {author} {\bibinfo {author} {\bibfnamefont {S.}~\bibnamefont
  {Van~der Jeught}}\ and\ \bibinfo {author} {\bibfnamefont {J.~J.}\
  \bibnamefont {Dirckx}},\ }\bibfield  {title} {\enquote {\bibinfo {title}
  {Deep neural networks for single shot structured light profilometry},}\
  }\href@noop {} {\bibfield  {journal} {\bibinfo  {journal} {Optics express}\
  }\textbf {\bibinfo {volume} {27}},\ \bibinfo {pages} {17091--17101} (\bibinfo
  {year} {2019})}\BibitemShut {NoStop}%
\bibitem [{\citenamefont {Yin}\ \emph {et~al.}(2019)\citenamefont {Yin},
  \citenamefont {Chen}, \citenamefont {Feng}, \citenamefont {Tao},
  \citenamefont {Huang}, \citenamefont {Trusiak}, \citenamefont {Asundi},\ and\
  \citenamefont {Zuo}}]{yin2019temporal}%
  \BibitemOpen
  \bibfield  {author} {\bibinfo {author} {\bibfnamefont {W.}~\bibnamefont
  {Yin}}, \bibinfo {author} {\bibfnamefont {Q.}~\bibnamefont {Chen}}, \bibinfo
  {author} {\bibfnamefont {S.}~\bibnamefont {Feng}}, \bibinfo {author}
  {\bibfnamefont {T.}~\bibnamefont {Tao}}, \bibinfo {author} {\bibfnamefont
  {L.}~\bibnamefont {Huang}}, \bibinfo {author} {\bibfnamefont
  {M.}~\bibnamefont {Trusiak}}, \bibinfo {author} {\bibfnamefont
  {A.}~\bibnamefont {Asundi}}, \ and\ \bibinfo {author} {\bibfnamefont
  {C.}~\bibnamefont {Zuo}},\ }\bibfield  {title} {\enquote {\bibinfo {title}
  {Temporal phase unwrapping using deep learning},}\ }\href@noop {} {\bibfield
  {journal} {\bibinfo  {journal} {Scientific Reports}\ }\textbf {\bibinfo
  {volume} {9}},\ \bibinfo {pages} {1--12} (\bibinfo {year}
  {2019})}\BibitemShut {NoStop}%
\bibitem [{\citenamefont {Hu}\ \emph {et~al.}(2019)\citenamefont {Hu},
  \citenamefont {Chen}, \citenamefont {Liang}, \citenamefont {Feng},
  \citenamefont {Tao},\ and\ \citenamefont {Zuo}}]{hu2019microscopic}%
  \BibitemOpen
  \bibfield  {author} {\bibinfo {author} {\bibfnamefont {Y.}~\bibnamefont
  {Hu}}, \bibinfo {author} {\bibfnamefont {Q.}~\bibnamefont {Chen}}, \bibinfo
  {author} {\bibfnamefont {Y.}~\bibnamefont {Liang}}, \bibinfo {author}
  {\bibfnamefont {S.}~\bibnamefont {Feng}}, \bibinfo {author} {\bibfnamefont
  {T.}~\bibnamefont {Tao}}, \ and\ \bibinfo {author} {\bibfnamefont
  {C.}~\bibnamefont {Zuo}},\ }\bibfield  {title} {\enquote {\bibinfo {title}
  {Microscopic 3d measurement of shiny surfaces based on a multi-frequency
  phase-shifting scheme},}\ }\href@noop {} {\bibfield  {journal} {\bibinfo
  {journal} {Optics and Lasers in Engineering}\ }\textbf {\bibinfo {volume}
  {122}},\ \bibinfo {pages} {1--7} (\bibinfo {year} {2019})}\BibitemShut
  {NoStop}%
\bibitem [{\citenamefont {An}, \citenamefont {Hyun},\ and\ \citenamefont
  {Zhang}(2016)}]{an2016pixel}%
  \BibitemOpen
  \bibfield  {author} {\bibinfo {author} {\bibfnamefont {Y.}~\bibnamefont
  {An}}, \bibinfo {author} {\bibfnamefont {J.-S.}\ \bibnamefont {Hyun}}, \ and\
  \bibinfo {author} {\bibfnamefont {S.}~\bibnamefont {Zhang}},\ }\bibfield
  {title} {\enquote {\bibinfo {title} {Pixel-wise absolute phase unwrapping
  using geometric constraints of structured light system},}\ }\href@noop {}
  {\bibfield  {journal} {\bibinfo  {journal} {Optics express}\ }\textbf
  {\bibinfo {volume} {24}},\ \bibinfo {pages} {18445--18459} (\bibinfo {year}
  {2016})}\BibitemShut {NoStop}%
\bibitem [{\citenamefont {Yin}\ \emph {et~al.}(2012)\citenamefont {Yin},
  \citenamefont {Peng}, \citenamefont {Li}, \citenamefont {Liu},\ and\
  \citenamefont {Gao}}]{yin2012calibration}%
  \BibitemOpen
  \bibfield  {author} {\bibinfo {author} {\bibfnamefont {Y.}~\bibnamefont
  {Yin}}, \bibinfo {author} {\bibfnamefont {X.}~\bibnamefont {Peng}}, \bibinfo
  {author} {\bibfnamefont {A.}~\bibnamefont {Li}}, \bibinfo {author}
  {\bibfnamefont {X.}~\bibnamefont {Liu}}, \ and\ \bibinfo {author}
  {\bibfnamefont {B.~Z.}\ \bibnamefont {Gao}},\ }\bibfield  {title} {\enquote
  {\bibinfo {title} {Calibration of fringe projection profilometry with bundle
  adjustment strategy},}\ }\href@noop {} {\bibfield  {journal} {\bibinfo
  {journal} {Optics letters}\ }\textbf {\bibinfo {volume} {37}},\ \bibinfo
  {pages} {542--544} (\bibinfo {year} {2012})}\BibitemShut {NoStop}%
\bibitem [{\citenamefont {He}\ \emph {et~al.}(2016)\citenamefont {He},
  \citenamefont {Zhang}, \citenamefont {Ren},\ and\ \citenamefont
  {Sun}}]{he2016deep}%
  \BibitemOpen
  \bibfield  {author} {\bibinfo {author} {\bibfnamefont {K.}~\bibnamefont
  {He}}, \bibinfo {author} {\bibfnamefont {X.}~\bibnamefont {Zhang}}, \bibinfo
  {author} {\bibfnamefont {S.}~\bibnamefont {Ren}}, \ and\ \bibinfo {author}
  {\bibfnamefont {J.}~\bibnamefont {Sun}},\ }\bibfield  {title} {\enquote
  {\bibinfo {title} {Deep residual learning for image recognition},}\ }in\
  \href@noop {} {\emph {\bibinfo {booktitle} {Proceedings of the IEEE
  conference on computer vision and pattern recognition}}}\ (\bibinfo {year}
  {2016})\ pp.\ \bibinfo {pages} {770--778}\BibitemShut {NoStop}%
\bibitem [{\citenamefont {Kingma}\ and\ \citenamefont
  {Ba}(2014)}]{kingma2014adam}%
  \BibitemOpen
  \bibfield  {author} {\bibinfo {author} {\bibfnamefont {D.~P.}\ \bibnamefont
  {Kingma}}\ and\ \bibinfo {author} {\bibfnamefont {J.}~\bibnamefont {Ba}},\
  }\bibfield  {title} {\enquote {\bibinfo {title} {Adam: A method for
  stochastic optimization},}\ }\href@noop {} {\bibfield  {journal} {\bibinfo
  {journal} {arXiv preprint arXiv:1412.6980}\ } (\bibinfo {year}
  {2014})}\BibitemShut {NoStop}%
\bibitem [{\citenamefont {Lohry}, \citenamefont {Chen},\ and\ \citenamefont
  {Zhang}(2014)}]{lohry2014absolute}%
  \BibitemOpen
  \bibfield  {author} {\bibinfo {author} {\bibfnamefont {W.}~\bibnamefont
  {Lohry}}, \bibinfo {author} {\bibfnamefont {V.}~\bibnamefont {Chen}}, \ and\
  \bibinfo {author} {\bibfnamefont {S.}~\bibnamefont {Zhang}},\ }\bibfield
  {title} {\enquote {\bibinfo {title} {Absolute three-dimensional shape
  measurement using coded fringe patterns without phase unwrapping or projector
  calibration},}\ }\href@noop {} {\bibfield  {journal} {\bibinfo  {journal}
  {Optics express}\ }\textbf {\bibinfo {volume} {22}},\ \bibinfo {pages}
  {1287--1301} (\bibinfo {year} {2014})}\BibitemShut {NoStop}%
\bibitem [{\citenamefont {Huang}, \citenamefont {Zhang},\ and\ \citenamefont
  {Asundi}(2013)}]{huang2013camera}%
  \BibitemOpen
  \bibfield  {author} {\bibinfo {author} {\bibfnamefont {L.}~\bibnamefont
  {Huang}}, \bibinfo {author} {\bibfnamefont {Q.}~\bibnamefont {Zhang}}, \ and\
  \bibinfo {author} {\bibfnamefont {A.}~\bibnamefont {Asundi}},\ }\bibfield
  {title} {\enquote {\bibinfo {title} {Camera calibration with active phase
  target: improvement on feature detection and optimization},}\ }\href@noop {}
  {\bibfield  {journal} {\bibinfo  {journal} {Optics letters}\ }\textbf
  {\bibinfo {volume} {38}},\ \bibinfo {pages} {1446--1448} (\bibinfo {year}
  {2013})}\BibitemShut {NoStop}%
\end{thebibliography}%


%

\end{document}